\documentclass[a4paper,11pt]{article}
\usepackage{pos}

\def\beq{\begin{equation}}  
\def\eeq{\end{equation}}
\def\({\left(}
\def\){\right)}
\def\[{\left[}
\def\]{\right]}

\title{Quasi-Gaussian Missing Higher Order Uncertainties}

\author*[a]{Marco Bonvini}
\author[b]{Emanuele Bagnaschi}
\author[c]{Lorenzo Paparella}

\affiliation[a]{INFN, Sezione di Roma,\\
  Piazzale Aldo Moro 5, Roma, Italy}

\affiliation[b]{INFN, Laboratori Nazionali di Frascati,\\
  Via E.~Fermi 54, Frascati, Italy}

\affiliation[c]{Department of Physics, Sapienza University,\\
  Piazzale Aldo Moro 5, Roma, Italy}

\emailAdd{marco.bonvini@roma1.infn.it}

\abstract{We review Bayesian models for estimating the uncertainty induced by missing
  higher orders in perturbative predictions.
  These models return probability distributions, which have peculiar shapes
  typically very different from the Gaussian distribution often assumed in physics analysis.
  In this work, we are able to obtain smooth probability distributions
  with quasi-Gaussians shapes by suitably modifying the likelihood and the priors of the models.
}

\FullConference{The 33rd International Workshop on Deep Inelastic Scattering and Related Subjects (DIS2026)\\
4 - 8 May 2026\\
Bologna, Italy\\}

\begin{document}
\maketitle

\section{Bayesian approaches to theory uncertainties from missing higher orders}

Perturbative predictions are necessarily truncated. For an observable
$\Sigma$, computed through N$^n$LO, the uncalculated remainder
$\Delta_{\rm MHO}=\Sigma-\Sigma_{\text{N$^n$LO}}$ is a source of theory
uncertainty which must be quantified whenever the prediction is used in a
precision analysis. The usual estimate is obtained by varying the
renormalisation and factorisation scales around a central value. Scale
dependence is formally of higher order, but the range of variation, the
central scale and the probabilistic interpretation of the resulting envelope
are conventional choices.

Bayesian inference offers a natural language for this
problem~\cite{Cacciari:2011ze, Bagnaschi:2014wea, Bonvini:2020xeo, Duhr:2021mfd}.
One specifies hypotheses for the behaviour of the
perturbative expansion, encoded in likelihoods and priors for a set of hidden
parameters. The known coefficients update these hypotheses and induce a
posterior distribution for the unknown coefficients, hence for
$\Delta_{\rm MHO}$. The result can be reported through credibility intervals,
but retaining the full distribution is useful when theory uncertainties must
be propagated or correlated.
An essential feature is that
the assumptions are explicit and can be tested, varied and improved.

\section{Geometric model and models based on scale dependence}

Following~\cite{Bonvini:2020xeo}, we write the generic perturbative series factoring out the LO term,
\begin{equation}
  \Sigma(\mu)=\Sigma_{\rm LO}(\mu)\sum_{k\geq0}\delta_k(\mu),
\end{equation}
so that the $\delta_k$ coefficients, which include the appropriate powers of $\alpha_s$,
are dimensionless numbers starting with $\delta_0=1$,
and $\mu$ is an unphysical scale (e.g., the renormalization scale).
The main models introduced in Ref.~\cite{Bonvini:2020xeo} are based on the conditions
\begin{subequations}\label{eq:modelcond}
\begin{align}
  |\delta_k(\mu)|&\leq c\,a^k, \label{eq:geo}\\
  |\delta_k(\mu)|&\leq\lambda\,r_{k-1}(\mu), \label{eq:sca}\\
  |r_k(\mu)|&\leq\eta\,r_{k-1}(\mu), \label{eq:cs}
\end{align}
\end{subequations}
where\footnote
{For the exact definition of $r_k$, which avoids problems
  in case of accidentally small scale dependences, see Ref.~\cite{Bonvini:2020xeo}.}
\begin{equation}
  r_k(\mu)\simeq\left|\mu\frac{d}{d\mu}
  \log\Sigma_{\rm N^kLO}(\mu)\right|
\end{equation}
represents a measure of the scale dependence at order $k$.
In Eq.~\eqref{eq:modelcond}, $c,a,\lambda,\eta$ are the parameters of the models,
for which one needs to provide a prior.
In particular, the \emph{geometric model} is based on Eq.~\eqref{eq:geo} and uses the prior
\begin{equation}
  \label{eq:geoprior}
  P_0(c,a) \propto \frac{\theta(1-c)}{c^{1+\epsilon}} \; (1-a)^\omega \theta(a)\,\theta(1-a),
\end{equation}
with $\epsilon,\omega$ parameters which can be varied to study the dependence of the model on the prior
(the suggested values are $\epsilon=0.1,\omega=1$).
The \emph{scale-variation model} is based on Eq.~\eqref{eq:sca} and uses the prior
\begin{equation}
  \label{eq:scaprior}
  P_0(\lambda) \propto \lambda^\gamma e^{-\lambda}\theta(\lambda),
\end{equation}
with $\gamma=1$ the suggested default value.
Finally, a variant of this model uses Eq.~\eqref{eq:sca} in conjunction with Eq.~\eqref{eq:cs},
which represent an additional constraint that gives less likelihood to higher orders
leading to stronger scale dependencies, with the additional prior
\begin{equation}
  \label{eq:csprior}
  P_0(\eta) \propto \eta^{\rho} e^{-\eta}\theta(\eta)
\end{equation}
and $\rho=0$ as default. We dub this the \emph{constrained scale-variation model}.
Note that it is possible to mix the conditions of Eq.~\eqref{eq:modelcond}
in different ways, even considering all of them at the same time.
The more conditions are considered, the more constrained the higher orders will be.

In all cases, the likelihood is chosen to strictly implement the respective condition in Eq.~\eqref{eq:modelcond},
which leads to the form
\begin{equation}
  \label{eq:likelihood}
  P(\delta_k|c,a) = \frac{\theta(ca^k-\delta_k)}{2ca^k},
\end{equation}
and similarly for the other models. The theta function implements the inequality,
and within the allowed range no preference is give to any value (flat distribution).

A characteristic of these models is to lead to distributions which are not Gaussians,
and in some cases they present features like spikes, plateaus, edges...
There is nothing wrong about these features. However, most physics analyses
assume Gaussian uncertainties, and upgrading these analyses to account for
a general distribution is far from trivial.

\section{Quasi Gaussian versions of the models}

Therefore, we now consider modification of the likelihood and the priors such that
the posterior distributions resemble a Gaussian distribution.\footnote
{Under reasonable conditions, it is impossible to produce exactly Gaussian distributions with these models at any order of the inference.
  I have a very nice proof of this, which this contribution is too short to contain.
}
Indeed, the actual form of likelihoods and priors is not fixed by the physics conditions Eq.~\eqref{eq:modelcond}.
Rather, when these conditions are translated into a probabilistic language there is some freedom.

We start from the likelihood. A generalization of Eq.~\eqref{eq:likelihood},
and similarly for the other models, is
\begin{equation}\label{eq:geolike}
P(\delta_k|c,a) = \frac N{ca^k}\, f\(\frac{|\delta_k|}{\beta ca^k}\)\, \theta\(\beta -\frac{|\delta_k|}{ca^k}\),
\end{equation}
where $N$ is a normalization factor and $f$ a function that can be used to favour some values
over others within the allowed range, which in turn is governed by $\beta\geq1$.
In addition to the flat likelihood of Eq.~\eqref{eq:likelihood}, we consider the variants:
\begin{itemize}
\item parabolic: $f(x) = 1-x^2$;
\item cosine: $f(x) = \cos^2\(\pi x/2\)$;
\item Gaussian: $f(x) = \exp\(-5x^2\)$.
\end{itemize}
In all cases, we have considered $\beta=1$, but higher values can be explored (they would lead to a broader allowed range,
typically producing larger uncertainties).
The parabolic and cosine likelihoods are constructed such that the function $f(x)$ goes to zero at $|x|=1$,
with the cosine one doing so in a smooth way, while the Gaussian one has truncated tails.
In all cases, the functional form is such that smaller higher orders (in absolute value) are preferred,
which may not be a good physical choice.

As far as the priors are concerned, we focus on those parameters with an infinite range, namely $c,\lambda,\eta$.
Indeed, these parameters are those that govern the tails of the posterior distributions~\cite{Bonvini:2020xeo}.
In order to obtain Gaussian-like tails, we need a strong (exponential) suppression.
The default prior used for the $c$ parameter in the geometric model, Eq.~\eqref{eq:geoprior},
is power-like, and as such is not suitable for this goal.
Therefore, we consider two variants:
\begin{itemize}
\item exponential: $P_0(c) \propto c^\kappa \exp(-c) \theta(c-1)$ (we used $\kappa=1$);
\item incomplete Gamma: $P_0(c) \propto \Gamma(s,c) \theta(c-1)$, with $s$ sufficiently large (we used $s=7$).
\end{itemize}
The incomplete Gamma prior can be adopted as a variant for the $\lambda,\eta$ parameters as well.

\begin{figure}[t]
 \centering
 \includegraphics[width=0.32\textwidth,page=1]{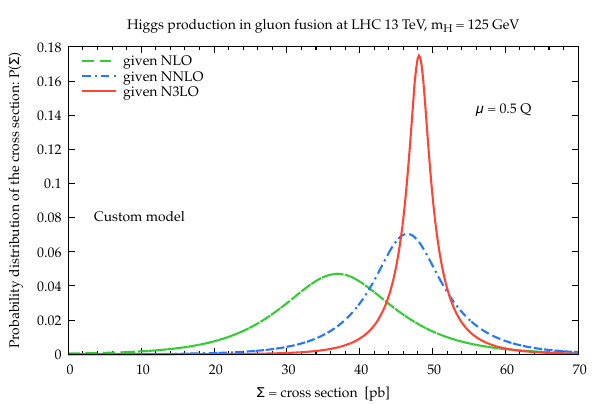}
 \includegraphics[width=0.32\textwidth,page=1]{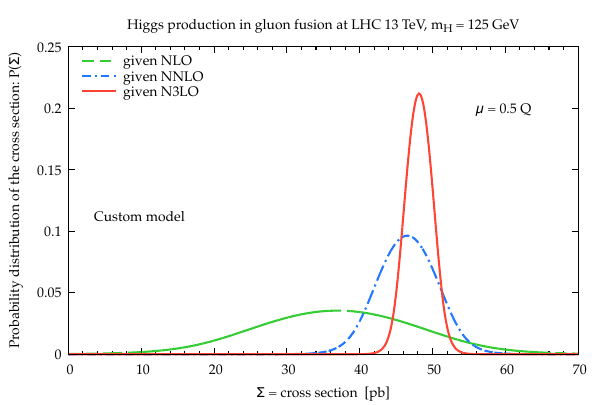}
 \includegraphics[width=0.32\textwidth,page=1]{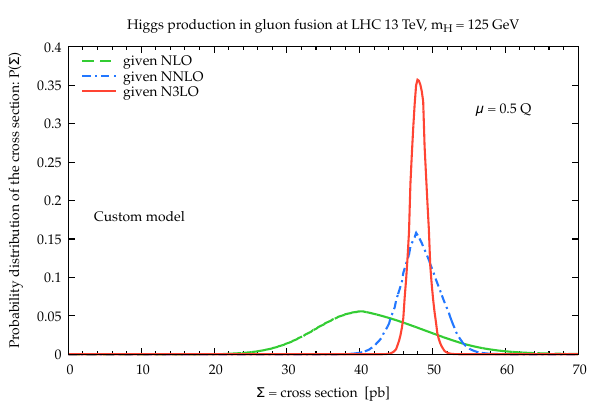}
 \caption{Probability distributions for the inclusive gluon-fusion Higgs
   cross section assuming the knowledge of the NLO (green), NNLO (blue) and N$^3$LO (red) result.
   The plots are obtained within the geometric (left), scale-variation (middle) and constrained scale-variation (right) models,
   using the modified likelihoods and prior as described in the text.}
 \label{fig:gauss}
\end{figure}

After testing all combinations, it turned out that the ``best'' shapes (namely those that look more like Gaussians)
are obtained by using the cosine likelihood and the exponential prior in all models considered.
The results for the three models applied to the case of Higgs production in gluon fusion
(the same prototypical example considered in Ref.~\cite{Bonvini:2020xeo}) are shown in figure~\ref{fig:gauss}.

In the case of the geometric model the distributions are smooth and with a bell shape
like a Gaussian, but a close inspection shows that the proportion are not compatible
with a Gaussian (taller peak and higher tails).

Instead, the scale-variation model with this combination of likelihood and prior
produces distributions that are very similar to Gaussians, as we could also verify by noting that
the standard deviation of the distribution is almost identical to the 68\% credibility interval,
which is in turn approximately half of the 95\% interval.

Finally, the constrained scale-variation model produces distributions that
are smoother than with the default likelihood, but which still present some spikes,
due to the nature of the constraint Eq.~\eqref{eq:cs}. Moreover, these distributions are asymmetric,
because the dependence of $r_k$ on $\delta_k$ is non linear. Nevertheless,
similarly to the scale-variation model, the statistical estimators approximately satisfy the same relations
of a Gaussian distribution, which means that it is possible to approximate these distribuitions
with Gaussians, obtaining a faithful representation of the features of this result.

\section{Conclusions}

Bayesian models for estimating missing higher order uncertainties have several advantages.
They allow to incorporate into a probabilistic framework any reasonable condition
that we believe a perturbative expansion must satisfy, making all the assumptions transparent and easily modifiable.
The models produce probability  distributions that can be propagated in phenomenological analyses.
However, simple incarnations of the models such as those presented in the original works
have peculiar shapes which are generally very different from the Gaussian distribution.

In this work, we have explored variants for likelihoods and priors to obtain
smoother, quasi-Gaussian distributions retaining the same physical principles while improving their
practical use in phenomenological analyses.
We have found that for models using scale dependence it is possible to achieve this goal,
with distributions that look Gaussians and satisfy the relations between standard deviation and credibility intervals
typical of Gaussians.
For the more general geometric model, the obtained distributions are smoother and with a Gaussian-like shape,
but not with the correct proportions. In this case, a Gaussian approximation may miss some qualitative features
of the model.

\end{document}